\documentclass[11pt]{article}
\begin{document}
\title
{Massive spin-$2$ and spin-$\displaystyle{{\frac12}}$ no hair
  theorems for stationary axisymmetric black holes} 
\author
{Sourav Bhattacharya{\footnote{souravbhatta@hri.res.in}}
\\
Harish-Chandra Research Institute, Chhatnag Road, Jhunsi, \\
Allahabad-211019, India,\\
and\\
Amitabha Lahiri, \footnote{amitabha@bose.res.in}\\
S. N. Bose National Centre for Basic Sciences, \\
Block JD, Sector III, Salt Lake, Kolkata -700098, India.\\
}

\maketitle
\abstract

{We present a proof of the no hair theorems corresponding to free
  massive non-perturbative Pauli-Fierz spin-$2$ and perturbative
  massive spin-$\frac12$ fields for stationary axisymmetric de
  Sitter black hole spacetimes of dimension four with two commuting
  Killing vector fields. The applicability of these results for
  asymptotically flat and anti-de Sitter spacetimes are also
  discussed. }

\hskip 1cm


\vskip 2cm

\section{Introduction}
The classical black hole no hair conjecture states that any
realistic gravitational collapse reaches a final stationary state
characterized by a small number of parameters. A part of this
conjecture has been proven mathematically rigorously by taking
different matter fields, known as the no hair theorem, (see
e.g. \cite{Chrusciel:1994sn, Heusler:1998ua, Heusler,
  Bekenstein:1998aw} and references therein) and deals with the
uniqueness of stationary black hole solutions characterized only by
mass, angular momentum, and charges corresponding only to long
range gauge fields. If a stationary black hole spacetime supports
in its exterior any non-trivial field configuration other than long
range gauge fields, the former one is called as `hair'. Thus
proving no hair theorems means to show that there cannot exist any
non-trivial and physically reasonable field configuration other
than long range gauge fields in the exterior of the black hole
spacetime. In particular, it has been shown that static,
spherically symmetric black hole spacetimes do not support hair
corresponding to scalars in convex potentials, Proca-massive vector
field~\cite{Bekenstein:1971hc}, or even gauge fields corresponding
to the Abelian Higgs model~\cite{Adler:1978dp, Lahiri:1993vg}.

However, all the above proofs assume asymptotic flatness, i.e. one
can reach spatial infinity and sufficiently rapid fall-off
conditions can be imposed upon the matter fields there. But recent
observations suggest that there is a strong possibility that our
universe is dominated by some exotic matter exerting negative
pressure such as a positive cosmological constant
$\Lambda$~\cite{Riess:1998cb, Perlmutter:1998np}. It is expected in
that case that the spacetime in its stationary state would possess
an outer or cosmological Killing
horizon~\cite{Bhattacharya:2010vr}. For known and exact stationary
solutions with a positive $\Lambda$~\cite{Carter:1968ks}, the
cosmological Killing horizon acts in general as a causal boundary
(see e.g. \cite{Gibbons:1977mu}) so that no observer can
communicate with region beyond this horizon along a future directed
path. If there is a black hole, the black hole event horizon will
be located inside the cosmological horizon and the spacetime is
then known as a de Sitter black hole spacetime. The observed value
of $\Lambda$ is tiny, of the order of $\displaystyle{ 10^{-52}
  {\rm{m}}^{-2}}$, and for such a small value the known solutions
show that the cosmological horizon has a length scale
$\displaystyle{\sim {\cal{O}}\left(\Lambda^{-\frac12}
  \right)}$. This is of course large, but not infinite. Since no
physical observer can communicate beyond the cosmological horizon,
in a de Sitter black hole spacetime the cosmological horizon serves
as a natural boundary along with the black hole horizon. So in
general one cannot impose any precise asymptotic fall off for the
matter fields in the vicinity of the cosmological horizon, nor can
set $T_{ab}=0$ there. Therefore, the generalization of the no hair
theorems for de Sitter black holes are expected to be different
from the $\Lambda \leq 0$ cases.

In fact considerable progress has been made in this topic for
static de Sitter black holes. Price's theorem, a perturbative no
hair theorem~\cite{Price:1971gc}, was proved
in~\cite{Chambers:1994sz} for massless perturbations in the
Schwarzschild-de Sitter background. Later the non-perturbative
black hole no hair theorems were extended for a general static de
Sitter black hole spacetime in~\cite{Bhattacharya:2007ap}. Notably
a violation of the standard no hair theorem was found -- a
spherically symmetric electrically charged solution sitting on the
false vacuum of the complex scalar of the Abelian Higgs model was
obtained which has no $\Lambda\leq 0$ analogue. In fact this
charged solution suggests that even though $\Lambda$ is tiny, the
existence of the cosmological horizon as an outer boundary of the
spacetime, because of the non-trivial boundary conditions, may
change local physics considerably. For some more aspects on no hair
theorems in such spacetimes we refer our reader to
\cite{Torii:1998ir, Martinez:2002ru}.

So it is an interesting task to generalize the no hair theorems
for stationary de Sitter black holes. For an asymptotically flat
spacetime, the no hair proofs for a rotating black hole for scalar
and Proca fields were given in~\cite{Bekenstein:1972ky}. The $\Lambda>0$ 
coordinate independent generalization of these proofs can be found in \cite{Bhattacharya:2011dq}.
For a discussion on the (2+1)-dimensional no hair theorem
see~\cite{Skakala:2009ss}. See also~\cite{Sen:1998bj} for a scalar
no hair theorem in stationary axisymmetric asymptotically flat spacetimes with
non-minimal matter-gravity coupling.  

In this paper we shall give a proof of the classical no hair
theorems corresponding to massive Pauli-Fierz spin-$2$~\cite{pauli}
and spin-$\frac12$ fields for stationary axisymmetric de Sitter
black hole spacetimes. For static asymptotically flat spherically
symmetric spacetime, a proof of spin-$2$ no hair can be found
in~\cite{Bekenstein:1972ky}.  It was shown later by constructing
Wu-Yang's magnetic monopole in such spacetimes that although
classical spin-$2$ hair is ruled out, quantum hair is not, which
can be detected via a stringy generalization of the Bohm-Aharonov
effect~\cite{Dvali:2006az, Dvali:2006nh}. We shall also address
briefly this phenomenon for these spacetimes. It was shown
in~\cite{Chambers:1994sz} that the Schwarzschild-de Sitter
spacetime does not support massless SL(2,~C) spinor hair with
vanishing frequency. For demonstration of the spin-$\frac12$ no
hair theorem via time dependent perturbation technique we refer our
reader to~\cite{Moderski:2008nq, Gibbons:2008gg, Gibbons:2008rs}.
We further refer our reader to e.g.~\cite{Shiromizu:2011he,
  Johannsen:2012ng, Rodriguez:2011aa, Johannsen:2011dh} and
references therein for recent developments including observational
aspects of the no hair theorem.

The paper is organized as follows. In the next section we outline
all the necessary assumptions and the geometrical set up we work
in. In Sec.s 3 and 4 we give respectively the proofs of the
classical no hair theorems for the massive spin-$2$ and
spin-$\frac12$ fields. Finally we discuss our results.

We shall set $c=G=\hbar=1$ throughout. We shall take mostly
negative signature $(+,~-,~-,~-)$ for the spacetime metric. For an
orthonormal basis $e_{(a)}^b$, the index in parenthesis will always
correspond to local Lorentz frame.

\section{Assumptions and the geometrical set up}
In the following we outline the assumptions and the geometrical set
up of the spacetime we work in, details of which can be found in
\cite{Bhattacharya:2010vr}.

The spacetime is a (3+1)-dimensional, smooth, connected,
orientable, Hausdorff and paracompact stationary axisymmetric
manifold with a Lorentzian metric $g_{ab}$, admits a spin
structure, satisfies Einstein's equations and is endowed with two
commuting Killing vector fields $\left\{\xi^a,~\phi^a\right\}$,
\begin{eqnarray}
\nabla_{(a}\xi_{b)} &=& 0 =
 \nabla_{(a}\phi_{b)} \,,\\
\left[\xi,~ \phi\right]^a &=&\pounds_{\xi}\phi^a=
\xi^b\nabla_b\phi^a-\phi^b\nabla_b\xi^a= 0\,. 
\label{g1}
\end{eqnarray}
$\xi^a$ is locally timelike with norm $\xi^a\xi_a = +\lambda^2$ and
generates the stationarity, whereas $\phi^a$ is locally spacelike
with closed orbits with parameter $0\leq \phi\leq 2\pi$ and norm
$\phi^a\phi_a = -f^2$ and hence generates the axisymmetry. We
assume that the spacetime connection `$\nabla$' is torsion free,
i.e. for any at least twice differentiable spacetime function
$\varepsilon(x)$ we have identically,
\begin{eqnarray}
\nabla_{[a}\nabla_{b]}\varepsilon(x) =0.
\label{g1'}
\end{eqnarray}
A basis for this spacetime can be chosen as
$\left\{\xi^a,~\phi^a,~\mu^a,~\nu^a\right\}$, where
$\left\{\mu^a,~\nu^a\right\}$ are spacelike basis vectors
orthogonal to both $\xi^a$ and $\phi^a$. We assume that the
spacelike 2-`planes' spanned by $\left\{\mu^a,~\nu^a\right\}$ form
integral submanifolds, i.e. $\mu^a$ and $\nu^a$ form the basis of a
Lie algebra.
%
%

For a stationary axisymmetric spacetime in general
$\xi^a\phi_a\neq0$, so the basis
$\left\{\xi^a,~\phi^a,~\mu^a,~\nu^a\right\}$ is not
orthogonal. Thus unlike static spacetimes, there exists no family
of spacelike hypersurfaces which is both tangent to $\phi^a$ and
orthogonal to $\xi^a$. Let us then first construct a family of
spacelike hypersurfaces tangent to $\phi^a$, which will be
convenient for our calculations. Let us define $\chi_{a}$ as
\begin{eqnarray}
\chi_a=\xi_a+\frac{1}{f^2}\left(\xi_b\phi^b\right)
\phi_a \equiv \xi_a+\alpha \phi_a,
\label{g2}
\end{eqnarray}
so that we have $\chi_a\phi^a=0$ everywhere. Also,
\begin{eqnarray}
\chi_a\chi^a  = \left(\lambda^2+\alpha^2 f^2\right)=\beta^2,
\label{g3}
\end{eqnarray}
so that $\chi_a$ is timelike when $\beta^2>0$. The basis
$\left\{\chi^a,~\phi^a,~\mu^a,~\nu^a\right\}$ thus serves as an
orthogonal basis for the spacetime. However, we note that $\chi^a$
is not a Killing field
\begin{eqnarray}
\nabla_{(a}\chi_{b)} = \phi_a\nabla_b \alpha
+\phi_b\nabla_a \alpha.
\label{g4}
\end{eqnarray}
We also note the following vanishing Lie derivatives which follow
immediately from Eq.s~(1) and (2), 
\begin{eqnarray}
\pounds_{\chi}\beta=0=\pounds_{\phi}\beta,\qquad
 \pounds_{\chi}\alpha=0=\pounds_{\phi}\alpha, \qquad
 \pounds_{\chi}f=0=\pounds_{\phi}f. 
\label{g5'nn}
\end{eqnarray}
In other words the 1-forms $\nabla_a\beta$, $\nabla_a\alpha$ and $\nabla_af$ 
are all orthogonal to both $\chi^a$ and $\phi^a$. This will prove
useful later.   

Our assumption that $\{\mu^a,~\nu^a\}$ span an integral
2-submanifold and Eq.s~(\ref{g5'nn}) imply that $\chi^a$ satisfies
the Frobenius condition of hypersurface
orthogonality~\cite{Bhattacharya:2010vr},
\begin{eqnarray}
\chi_{[a}\nabla_b\chi_{c]}=0.
\label{g5}
\end{eqnarray}
Thus $\chi^a$ is orthogonal to the spacelike
$\left\{\phi^a,~\mu^a,~\nu^a\right\}$ hypersurfaces, say
$\Sigma$. Using Eq.s~(\ref{g4}) and (\ref{g5}), we get an useful
expression
\begin{eqnarray}
\nabla_a\chi_{b}=\beta^{-1}\chi_{[b}\nabla_{a]}\beta +
\frac12\phi_{(a}\nabla_{b)}\alpha.  
\label{g5'n}
\end{eqnarray}
We are dealing with a stationary axisymmetric spacetime with two
Killing horizons. One is the black hole horizon and the larger one
which surrounds the black hole is the cosmological horizon.  Let us
now locate the horizons in terms of the orthogonal basis
$\left\{\chi^a,~\phi^a,~\mu^a, ~\nu^a\right\}$. A stationary
axisymmetric spacetime with a black hole is in general rotating and
in that case $\xi^a$ becomes spacelike within the
ergosphere~\cite{Wald:1984rg}, so for such spacetimes the surface
$\lambda^2=0$ does not in general define a horizon.  It was shown
in~\cite{Bhattacharya:2010vr} by considering the null geodesic
congruence tangent to a `closed' $\beta^2=0$ hypersurface
${\cal{H}}$ that the function $\alpha$ is a constant on ${\cal{H}}$
and the orthogonal vector field $\chi^a$ coincides with a null
Killing field there. Thus any such surface ${\cal{H}}$ is
essentially a Killing or true horizon. Accordingly, we define the
black hole and the cosmological event horizons to be the two
`closed' $\beta^2=0$ surfaces, the former being located inside the
second, such that $\chi^a$ is timelike in the region between them,
becoming null on the surfaces.  An example of this is the
Kerr-Newman-de Sitter family of spacetimes \cite{Gibbons:1977mu}.

We note that there could be a Cauchy horizon too, located inside
the black hole event horizon. This is another closed $\beta^2=0$
surface on which $\chi^a$ is Killing and null, however the vector
field $\chi^a$ is spacelike between this surface and the event
horizon.
The existence of the Cauchy horizon makes the black hole
singularity timelike, resulting in interesting consequences in
analytically extended
charts~\cite{Chandrasekhar:1985kt}. Perturbative studies show that
the Cauchy horizon can be unstable. We refer our readers
to~\cite{Chandrasekhar:1985kt} (also references therein) for an an
excellent account on this for $\Lambda=0$.  For $\Lambda>0$, this
result was generalized later in~\cite{Chambers:1994ap}. We further
refer our reader to~\cite{Gibbons:1977mu} for maximal analytic
extension of the Kerr-Newman-de Sitter spacetime including the
Cauchy horizon. However it is sufficient for our present purpose to
consider only the region between the black hole event horizon and
the cosmological horizon, and we can safely ignore the inner Cauchy
horizon if it exists.

For convenience of our calculation, we shall specify $\mu^a$
now. On any of ${\cal{H}}$ we know that 
\begin{eqnarray}
\nabla_a\beta^2=2\kappa\chi_a,
\label{g6'}
\end{eqnarray}
where $\kappa$ is a constant on ${\cal{H}}$ known as the respective
surface gravity. 
 Keeping in mind that $\nabla_a\beta^2$
is orthogonal to both $\chi^a$ and $\phi^a$ (Eq.s~(\ref{g5'nn})),
we define  
\begin{eqnarray}
\mu_a:=\frac{1}{2\kappa(x)}\nabla_a\beta^2,
\label{g6''}
\end{eqnarray}
where $\kappa(x)$ is a function which smoothly reaches $\kappa$
when we reach ${\cal{H}}$. With this choice $\mu^a$ is itself
Killing and null on ${\cal{H}}$ and vanishes there as
${\cal{O}}(\beta^2)$. When the black hole is extremal, i.e.
$\kappa=0$, we simply write $\mu^a = \frac12\nabla_a\beta^2\,.$

The projector $h_{a}{}^b$ which projects tensors onto the spacelike
hypersurfaces $\Sigma$ is defined as 
\begin{eqnarray}
h_{a}{}^{b}=\delta_{a}{}^{b}-\beta^{-2}\chi_a\chi^b.
\label{g6}
\end{eqnarray}
Let $D_a$ be the spacelike induced connection defined via the
projector as $D_a\equiv h_{a}{}^{b}\nabla_b$. Then we can project
the derivative of a tensor $T_{a_1a_2\cdots}{}^{b_1b_2\cdots}$ onto
$\Sigma$ as
\begin{eqnarray}
D_a\widetilde{T}_{a_1a_2\dots}{}^{b_1b_2\dots}:=h_{a}{}^{b}
 h_{a_1}{}^{c_1}\dots h^{b_1}{}_{d_1}\dots\nabla_b
T_{c_1c_2\dots}{}^{d_1d_2\dots},
\label{g7}
\end{eqnarray}
where $\widetilde{T}$ is the projection of $T$ onto $\Sigma,$ given
by $\widetilde{T}_{a_1a_2\cdots}{}^{b_1b_2\cdots} :=
h_{a_1}{}^{c_1}\cdots h^{b_1}{}_{d_1}\cdots
T_{c_1c_2\cdots}{}^{d_1d_2\cdots}$. It is easy to verify that the
induced connection $D_a$ on $\Sigma$ defined in Eq.~(\ref{g7})
satisfies the Leibniz rule and is compatible with the induced
metric $h_{ab}$.  For our purpose we shall also need to act `$D$'
on a full spacetime tensor $T$ by
\begin{eqnarray}
D_aT_{a_1a_2\dots}{}^{b_1b_2\dots}:=h_{a}{}^{b}\nabla_b
T_{a_1a_2\dots}{}^{b_1b_2\dots},
\label{g7.1}
\end{eqnarray}
in which it is clear that $D_a$ is merely the spacelike directional
derivative associated with the full metric.

We shall also need to project tensors onto the integral 2-planes
orthogonal to both $\chi^a$ and $\phi^a$ and spanned by $\mu^a$ and
$\nu^a$, say $\overline{\Sigma}$. The projection tensor is given by
\begin{eqnarray}
\pi_{a}{}^{b}=\delta_{a}{}^{b}-\beta^{-2}\chi_a\chi^b+f^{-2}\phi_a\phi^b. 
\label{g8}
\end{eqnarray}
The projected derivative `$\overline{D}$' on $\overline{\Sigma}$
can be defined exactly in the same way as above.

Using the fact that the 2-planes spanned by $\mu^a$ and $\nu^a$ are
integral submanifolds, we can derive the following expression for
the derivative of the Killing field
$\phi^a$~\cite{Bhattacharya:2010vr},
\begin{eqnarray}
\nabla_a\phi_{b}=f^{-1}\phi_{[b}\nabla_{a]}f+\frac12\chi_{[a}\nabla_{b]}\alpha.
\label{g9''}
\end{eqnarray}
We assume that there is no naked curvature singularity anywhere in
our region of interest, i.e. anywhere between the two horizons
including both of them. The Einstein equation $G_{ab}+\Lambda
g_{ab}=8 \pi T_{ab}$ then implies that the invariants constructed
from the energy-momentum tensor $T_{ab}$ are bounded everywhere in
our region of interest.

We assume that any physical matter field, or any observable
concerning the matter field also obeys the symmetries of the
spacetime, be it continuous or discrete, because otherwise the
matter field may itself break those symmetries. In other words, if
$X$ is a physical matter field or a component of it, or an
observable quantity associated with it, we must have
\begin{eqnarray}
\pounds_{\xi}X=0=\pounds_{\phi}X.
\label{sym}
\end{eqnarray}
Apart from the existence of the cosmological horizon as an outer
boundary and regularity, no asymptotics on spacetime or matter
fields will be imposed. However unlike the spin-$2$ field, we shall
ignore backreaction of the spinor on the spacetime since spinors do
not obey any classical energy condition~\cite{Penrose:1985jw}. We
shall not consider any coupling of the spinor with gauge fields.
We shall not explicitly solve Einstein's equations but shall only
examine the existence of solutions of matter fields.
 
Being equipped with all this, we are now ready to go into the no
hair proofs.

\section{Massive spin-$2$ field} 
Let us begin with the massive and real spin-$2$ field $M_{ab}$. An
equation of motion for $M_{ab}$ can be written
as~\cite{Bekenstein:1972ky, pauli}
\begin{eqnarray}
\nabla_c\nabla^c \left( M_{ab}-\frac12Mg_{ab}\right)+m^2
\left(M_{ab}-\frac12Mg_{ab}\right)=0. 
\label{nh2}
\end{eqnarray}
$M_{ab}$ is symmetric in its two indices, $M=M_{ab}g^{ab}$ and $m$
can be interpreted as the rest mass of the field. $M_{ab}$
satisfies the condition : $\nabla_a M^{a}{}_b=0$. We note here that
unlike the gravitational perturbation equation, a pure spin-$2$
field theory has some ambiguities in its coupling with spacetime
curvature. In particular, Eq.~(\ref{nh2}) might have contained
terms like $R_{acbd}M^{cd}$. However, under the reasonable
assumption that the Compton wavelength of the field is small
compared to the size of the black hole horizon, the mass term
always dominates over such terms outside the
horizon~\cite{Bekenstein:1972ky}. So, we shall not consider
non-minimal coupling of the field with curvature.

We take the trace of Eq.~(\ref{nh2}) and note that since $M$ is a
scalar, $\pounds_{\chi}M=
\pounds_{\xi}M+\alpha\pounds_{\phi}M=0$. Using this and
Eq.~(\ref{g6}), we find
\begin{eqnarray}
\nabla_a \nabla^aM=\frac{1}{\beta h}\partial_{c}\left[\beta
  hg^{cd}\partial_d M\right]=\frac{1}{\beta
  h}\partial_{c}\left[\beta hh^{cd}\partial_d
  M\right]=\frac{1}{\beta}D_a\left(\beta D^aM\right), 
\label{nhad1}
\end{eqnarray}
where $h$ is the determinant of the induced metric $h_{ab}$. Thus
the trace of Eq.~(\ref{nh2}) is equivalent to 
\begin{eqnarray}
D_a\left(\beta D^aM\right)+m^2\beta M=0, 
\label{nhad2}
\end{eqnarray}
which we multiply with $M$ and integrate by parts on $\Sigma$
between the two horizons. The total divergence term is converted to
a surface integral on ${\cal{H}}~(\beta=0)$ and goes away leaving
with us the vanishing volume integral,
\begin{eqnarray}
\int_{\Sigma}\beta\left[-\left(D_aM\right)\left(D^aM\right)+m^2
  M^2\right]=0,  
\label{nhad3}
\end{eqnarray}
which shows $M=0$ throughout.

In four spacetime dimensions $M_{ab}$ has ten components,
\begin{eqnarray}
M_{ab}=\Psi^{(1)}\chi_a\chi_b+\Psi^{(2)}\phi_a\phi_b +
\Psi^{(3)}\mu_a\mu_b+\Psi^{(4)}\nu_a\nu_b + 
\Psi^{(5)}\chi_{(a}\phi_{b)}+
\Psi^{(6)}\chi_{(a}\mu_{b)}+\nonumber\\
\Psi^{(7)}\chi_{(a}\nu_{b)}+\Psi^{(8)}\phi_{(a}\mu_{b)} +
\Psi^{(9)}\phi_{(a}\nu_{b)}+ 
\Psi^{(10)}\mu_{(a}\nu_{b)},
\label{nhad4}
\end{eqnarray}
where $\Psi^{(i)}$'s are scalars. To simplify our calculations, we
shall now use the discrete symmetry of the spacetime to get rid of
some of these components of $M_{ab}$.  The metric for a stationary
axisymmetric spacetime under consideration is invariant under the
simultaneous reflections $\xi_a\to -\xi_a$ and
$\phi_a\to-\phi_a$. Eq.~(\ref{g2}) then shows these are equivalent
to $\chi_a\to-\chi_a$ and $\phi_a\to -\phi_a$. Since we are not
ignoring backreaction, any physical matter field must obey these
symmetries \cite{Bekenstein:1972ky, weinberg}. Noting that all the
scalars in Eq.~(\ref{nhad4}) are independent of parameters along
$\xi^a$ and $\phi^a$, we find that the invariance under the
discrete symmetry implies $\Psi^{(6)}=\Psi^{(7)}=
\Psi^{(8)}=\Psi^{(9)}=0$.

Thus we are left with six components of $M_{ab}$ :
$\left\{\chi\chi,~\phi\phi,~\chi\phi,~\mu\mu,
  ~\nu\nu,~\mu\nu\right\}$. For simplicity of notation, we shall
denote the orthogonal directions $(\chi,~\phi,~\mu,~\nu)$ as
$(0,~1,~2,~3)$ respectively.

Since $M_{ab}$ is a physical matter field, by Eq.~(\ref{sym}) we
have $\pounds_{\xi}M_{ab}=0=\pounds_{\phi}M_{ab}$. This gives
\begin{eqnarray}
\pounds_{\chi}M_{ab}=\chi^c\nabla_c
M_{ab}+M_{cb}\nabla_a\chi^c+M_{ca}\nabla_b\chi^c =
\phi^cM_{c(b}\nabla_{a)}\alpha.  
\label{nh3''}
\end{eqnarray}
Using Eq.~(\ref{g5'n}), we find from the above equation 
\begin{eqnarray}
\chi^c\nabla_c
M_{ab}=\frac12\phi^cM_{c(b}\nabla_{a)}
\alpha-\frac12\left(\nabla^c\alpha\right)
M_{c(b}\phi_{a)}  
+\beta^{-1}\left(\nabla^c\beta\right)M_{c(b}\chi_{a)} -
\beta^{-1}\chi^cM_{c(b}\nabla_{a)}\beta 
\nonumber\\=H_{ab}~(\rm say).
\label{nh3''n}
\end{eqnarray}
Using this and the fact that $M=0$ we now find from
Eq.~(\ref{nh2}), 
\begin{eqnarray}
\int_{{\cal{H}}}M_{ab}\nabla_cM_{ab}d{\cal{H}}^c+\int [dX]
\left[-\beta^{-2}H_{ab}^2 
-\left(D_c M_{ab} \right)\left(D^cM_{ab}\right)+m^2M_{ab}^2 \right]
~({\rm{no~sum~on}}~a,~b), 
\label{nhad5}
\end{eqnarray}
where $[dX]$ is the full spacetime volume measure and the direction
`$c$' in the horizon integral directs along $\mu^a$. By our choice
$\mu^a$ coincides with $\chi^a$ on ${\cal{H}}$ (Eq.s
~(\ref{g6'}),~(\ref{g6''})), so that the integrand in the horizon
integral coincides with $M_{ab}H_{ab}$. Let us first set $a=0$,
$b=1$ in the above integrals. Using the fact that $\nabla_a\alpha$
and $\nabla_a\beta$ are both orthogonal to $\chi^a$ and $\phi^a$
(Eq.s~(\ref{g5'nn})), and four of the ten components of $M_{ab}$
are already zero, we find from Eq.~(\ref{nh3''n}) that
$H_{01}=0$. Then Eq.~(\ref{nhad5}) shows that $M_{01}=0$
throughout. Similarly we can show that all the other components of
$M_{ab}$ vanish also.

Thus all the six components of $M_{ab}$ vanish identically in the
region between the black hole and cosmological horizon. This is the
expected classical no hair result for this field. For
asymptotically flat or anti-de Sitter spacetimes ($\Lambda\leq0$),
the boundary integral at the cosmological horizon is replaced by an
integral at spacelike infinity. By imposing sufficiently rapid
fall-off condition on the matter field, we can make the integral
vanishing and the desired no hair result follows.

It was shown in~\cite{Dvali:2006az, Dvali:2006nh} for static
spherically symmetric spacetimes that although classical spin-$2$
hair is ruled out, quantum hair is not.  The idea is the
following. A St\"{u}ckelberg field $A_b$ was introduced to write
$M_{ab}$ as
\begin{eqnarray}
M_{ab}=\widehat{M}_{ab}+\nabla_aA_b+\nabla_bA_a.
\label{qh1}
\end{eqnarray}
Then $M_{ab}$ is invariant under the local gauge transformations :
$A_b\to A_b-\zeta_b, ~\widehat{M}_{ab}\to
\widehat{M}_{ab}+\nabla_{(a}\zeta_{b)}$.
Since $M_{ab}=0$, one has
$\widehat{M}_{ab}=-\left(\nabla_aA_b+\nabla_bA_a\right)$. Then a
magnetic monopole solution for $F_{ab}=\nabla_{[a}A_{b]}$ was
constructed and it was shown that the magnetic charge can be
detected via a stringy generalization of the Bohm-Aharonov effect
in the asymptotic region. In this work we have shown that $M_{ab}$
vanishes also for general stationary axisymmetric
spacetimes. Following this, we can break $M_{ab}$ into two gauge
fields, from one of which we can construct a magnetic monopole
solution.  It is clear that the solution will not be spherically
symmetric in this case.  However, if the black hole is small
compared to the cosmological horizon size, spacetime will be
spherically symmetric at large distance from the black hole, and
the solution will asymptotically reach the usual spherically
symmetric monopole solution. Accordingly, we can detect in this
region a magnetic charge of the black hole.  It remains as an
interesting task to construct explicitly such monopole solutions,
for example for the Kerr-de Sitter spacetime.

\section{Massive spin-$\frac12$ field} 
Let us now consider the case of a massive spin-$\frac12$ field. The
detailed formalism of such fields in curved spacetime can be found
in e.g.~\cite{Penrose:1985jw, Wald:1984rg, Parker}.  The Lagrangian
is given by
\begin{eqnarray}
{\cal{L}}=\frac{i}{2}\left[\overline{\Psi}\gamma^a\nabla_a\Psi -
  \left(\nabla_a \overline{\Psi}\right)\gamma^a\Psi\right] -
m\overline{\Psi}\Psi,   
\label{spin1}
\end{eqnarray}
where $\Psi$ is a 4-component spinor. The covariantly constant
matrices $\gamma^a$'s can be expanded in an orthonormal basis
$\gamma^a=e^a_{(b)}\gamma^{(b)}$.  Using the well known
anticommutation relation,
$[\gamma^{(a)},~\gamma^{(b)}]_+=2\eta^{(a)(b)}\bf{I}$, where
$\bf{I}$ is the $4\times4$ identity matrix, we find
\begin{eqnarray}
[\gamma^a,~\gamma^b]_+=2g^{ab}\bf{I}.
\label{spin2}
\end{eqnarray}
The adjoint spinor $\overline{\Psi}$ is defined as
$\overline{\Psi}=\Psi^{\dagger}\gamma^{(0)}$.  The matrix
$\gamma^{(0)}$ is Hermitian whereas $\gamma^{(i)}$, $i=1,~2,~3,$
are anti-Hermitian.  The spin covariant derivative `$\nabla$' in
Eq.~(\ref{spin1}) is defined as
\begin{eqnarray}
\nabla_a\Psi=\partial_a\Psi+\frac18\omega_{a(b)(c)}
[\gamma^{(b)},~\gamma^{(c)}]\Psi, 
\quad \nabla_a 
\overline{\Psi}=\partial_a\overline{\Psi}-
\frac{\overline{\Psi}}{8}\omega_{a(b)(c)}[\gamma^{(b)},~\gamma^{(c)}],
\label{spin3}
\end{eqnarray}
where $\omega_{a(b)(c)}$ are the Ricci rotation coefficients given
by $\omega_{a(b)(c)}=e_{(b)}^d\nabla_a e_{(c)d}$. It is easy to
show using Eq.~(\ref{spin3}) that~\cite{Penrose:1985jw,
  Wald:1984rg, Parker},
\begin{eqnarray}
[\nabla_a,~\nabla_b]\Psi=-\frac18R_{ab(c)(d)}
[\gamma^{(c)},~\gamma^{(d)}]\Psi=-\frac18R_{abcd}
[\gamma^{c},~\gamma^{d}]\Psi,   
\label{spin4}
\end{eqnarray}
using the fact that contraction is independent of basis. The
equations of motion are given by
\begin{eqnarray}
i\gamma^a\nabla_a\Psi-m\Psi=0,\quad
i\left(\nabla_{a}\overline{\Psi}\right)\gamma^a+ 
m\overline{\Psi}=0.
\label{spin5}
\end{eqnarray}
We consider the conserved current 1-form $J_a$,
\begin{eqnarray}
J_a=\overline{\Psi}\gamma_a\Psi~:~\nabla_aJ^a=0, 
\label{add1}
\end{eqnarray}
by Eq.s~(\ref{spin5}). Let us define a 2-form $S_{ab}$, 
\begin{eqnarray}
S_{ab}:=\nabla_{[a}J_{b]},
\label{add2}
\end{eqnarray}
so that
\begin{eqnarray}
\nabla^aS_{ab}=\nabla_a\nabla^a\left(\overline{\Psi}
  \gamma_b\Psi\right)-R_{b}{}^{a}
\left(\overline{\Psi}\gamma_a\Psi\right).   
\label{add3}
\end{eqnarray}
Since we ignore backreaction in this case, we have $R_{ab}=\Lambda
g_{ab}$. Then setting $b=0$ above and noting
$\gamma_0=\beta\gamma_{(0)}$ in our orthogonal basis, we find the
following
\begin{eqnarray}
\nabla^aS_{a0}=\nabla_a\nabla^a\left(\beta\Psi^{\dagger}\Psi\right)
-\Lambda\beta\Psi^{\dagger}\Psi.
\label{add4}
\end{eqnarray}
Integrating the above equation using the full spacetime volume
element $[dX]$ and converting the total 
divergences into surface integrals on ${\cal{H}}$ we get
\begin{eqnarray}
\int_{{\cal{H}}}S_{a0}d{\cal{H}}^a
-\int_{{\cal{H}}}\nabla_a\left(\beta\Psi^{\dagger}\Psi\right) 
d{\cal{H}}^a
+\int[dX]\Lambda\beta \Psi^{\dagger} \Psi=0,
\label{add5}
\end{eqnarray}
where the unit normal `$a$' as before directs along $\mu^a$. It is
clear that the measures on ${\cal{H}}$ are non-divergent. Since
$S_{ab}$ is antisymmetric in its indices and by our choice $\mu_a$
coincides with $\chi_a$ on ${\cal{H}}$ (Eq.s~(\ref{g6'}),
(\ref{g6''})), the first integral vanishes in Eq.~(\ref{add5}). Let
us now evaluate the second boundary integral. Eq.~(\ref{sym})
implies $\pounds_{\xi}J_a=0=\pounds_{\phi}J_a$, which gives
\begin{eqnarray}
\chi^a\nabla_a\left(\overline{\Psi}\gamma_b\Psi\right)=
\frac{\left(\overline{\Psi}\gamma^a\Psi\right)}{2\beta^2}
\left[\chi_a\nabla_b\beta^2-\chi_b\nabla_a\beta^2
\right]+\frac12\left(\overline{\Psi}\gamma^a\Psi\right) 
\left[\phi_a\nabla_b\alpha-\phi_b\nabla_a\alpha\right],
\label{add6}
\end{eqnarray}
where we have used Eq.~(\ref{g5'n}). Setting $b=0$ and using
Eq.s~(\ref{g5'nn}) we get 
\begin{eqnarray}
\chi^a\nabla_a\left(\beta\Psi^{\dagger}\Psi\right)=
-\frac{\left(\overline{\Psi}
    \gamma_a\Psi\nabla^a\beta^2\right)}{2\beta^2}\chi_0.   
\label{add7} 
\end{eqnarray}
Since $\mu^a$ coincides with $\chi^a$ on ${\cal{H}}$, the second
integrand in Eq.~(\ref{add5}) is given by the above expression.
Then from the fact that
$\overline{\Psi}\gamma_0\Psi=\beta\Psi^{\dagger}\Psi$, it is clear
that the above quantity is ${\cal{O}}(\beta)$ when evaluated on
${\cal{H}}$. This implies the second integral in Eq.~({\ref{add5}})
also vanishes. This shows that $\Psi=0$ throughout. For $\Lambda<0$
the outer boundary is infinity and suitable fall-off condition for
the massive field recovers the no hair result.

The above simple proof is however not valid for
$\Lambda=0$. Unfortunately we have been able to do the proof for
such spacetimes only under stronger assumption than the above. It
is the following.

We multiply the first of Eq.s~(\ref{spin5}) by $i\gamma^b\nabla_b$
and use Eq.s~(\ref{spin2}), (\ref{spin4}) to get
\begin{eqnarray}
\nabla_a\nabla^a\Psi-\frac{1}{32}R_{abcd}
[\gamma^{a},~\gamma^{b}][\gamma^{c},~\gamma^{d}]\Psi  
+m^2\Psi=0.
\label{spin6}
\end{eqnarray}
We shall now simplify the second term. Denoting
$[\gamma^{a},~\gamma^{b}]$ by $\widetilde{\sigma}^{ab}$, we compute
\begin{eqnarray}
\widetilde{\sigma}^{a[b}\widetilde{\sigma}^{cd]} =
2\left[\widetilde{\sigma}^{ab}\widetilde{\sigma}^{cd} +
  \widetilde{\sigma}^{ac}\widetilde{\sigma}^{db} +
  \widetilde{\sigma}^{ad}\widetilde{\sigma}^{bc}\right].  
\label{spin7}
\end{eqnarray}
Contracting both sides by $R_{abcd}$, recalling the identity
$R_{a[bcd]}=0$, and the symmetries of the Riemann tensor we find 
\begin{eqnarray}
R_{abcd}\widetilde{\sigma}^{ab}\widetilde{\sigma}^{cd} =
2R_{abcd}\widetilde{\sigma}^{ac}\widetilde{\sigma}^{bd}. 
\label{spin8}
\end{eqnarray}
Using the anticommutation relations for the $\gamma$'s we find from
the above
\begin{eqnarray}
R_{abcd}\widetilde{\sigma}^{ab}\widetilde{\sigma}^{cd} =
8\left[R_{abcd}\gamma^a\gamma^c\gamma^b\gamma^d 
-R\right].
\label{spin9}
\end{eqnarray}
The first term can be written as 
\begin{eqnarray}
R_{abcd}\gamma^a\gamma^c\gamma^b\gamma^d =
R_{abcd}\gamma^a\left(2g^{bc}-\gamma^b\gamma^c\right)\gamma^d 
=-2R-R_{abcd}\gamma^a\gamma^b\gamma^c\gamma^d =
-2R-\frac14R_{abcd}\widetilde{\sigma}^{ab}\widetilde{\sigma}^{cd}, 
\label{spin12}
\end{eqnarray}
using the fact that
$R_{abcd}\left(\gamma^a\gamma^b-\gamma^b\gamma^a\right)\gamma^c\gamma^d=
R_{abcd}\left(2\gamma^a\gamma^b-2g^{ab}\right)\gamma^c\gamma^d=
2R_{abcd}\gamma^a\gamma^b\gamma^c\gamma^d$. Putting in all this we
have from Eq.~(\ref{spin9})
\begin{eqnarray}
R_{abcd}\widetilde{\sigma}^{ab}\widetilde{\sigma}^{cd}=-8R.
\label{spin13}
\end{eqnarray}
Thus Eq.~(\ref{spin6}) now simplifies to~\cite{Parker},
\begin{eqnarray}
\nabla_a\nabla^a\Psi+\left(\frac{R}{4}+m^2\right)\Psi=0.
\label{spin14}
\end{eqnarray}
It can be seen from Eq.~(\ref{spin3}) that
$\left(\nabla_a\Psi\right)^{\dagger}= \nabla_a\Psi^{\dagger}$, so
that $\Psi^{\dagger}$ satisfies the same equation as above.  From
Einstein's equations we get
\begin{eqnarray}
R=4\left(\Lambda -2\pi T\right).
\label{spin15}
\end{eqnarray}
%
%
%
Multiplying Eq.~(\ref{spin14}) by $\Psi^{\dagger}$ and using the
projector $\pi_{a}{}^{b}$ defined in Eq.~(\ref{g9''}) we now
compute
\begin{eqnarray}
  \overline{D}_a\overline{D}^a\left(\Psi^{\dagger}\Psi\right)
  :=\pi^{ab}\nabla_a\nabla_{b} 
  \left(\Psi^{\dagger}\Psi\right)=2\left(\nabla_a\Psi^{\dagger}\right)
  \left(\nabla_a\Psi\right)-2\left(m^2+\frac{R}{4}\right)
  \Psi^{\dagger}\Psi \nonumber\\+ 
  \left[f^{-2}\phi^a\nabla_a\left(\phi^b\nabla_b
      \left(\Psi^{\dagger}\Psi\right) \right)  
    -\beta^{-2}\chi^a\nabla_a\left(\chi^b\nabla_b
      \left(\Psi^{\dagger}\Psi\right) \right) 
  \right]\nonumber\\-\left[f^{-1}\left(\overline{D}_af\right)
    \left(\overline{D}_a
      \left(\Psi^{\dagger}\Psi\right)\right)+\beta^{-1}
    \left(\overline{D}_a\beta\right)\left(\overline{D}_a
      \left(\Psi^{\dagger}\Psi\right)\right)\right], 
\label{spin17}
\end{eqnarray}
where we have used equations of motion for $\Psi$ and
$\Psi^{\dagger}$ and the fact that
$\nabla_a\beta=D_a\beta=\overline{D}_a\beta$ and
$\nabla_af=D_af=\overline{D}_af$, which follow from
Eq.s~(\ref{g5'nn}). The above can be rewritten as
\begin{eqnarray}
\overline{D}_a\left[f\beta\overline{D}^a
  \left(\Psi^{\dagger}\Psi\right)\right]=2f\beta 
\left(\overline{D}_a\Psi^{\dagger}\right)
\left(\overline{D}_a\Psi\right)-
2f\beta\left(m^2+\frac{R}{4}\right)\Psi^{\dagger}\Psi \nonumber\\ 
+
f\beta\left[f^{-2}\left[\phi^a\nabla_a\left(\phi^b\nabla_b
      \Psi^{\dagger}\right)\Psi+ 
\Psi^{\dagger}\phi^a\nabla_a\left(\phi^b\nabla_b \Psi\right)\right]
\right.\nonumber\\
\left.-\beta^{-2}\left[\chi^a\nabla_a\left(\chi^b\nabla_b
      \Psi^{\dagger}\right)\Psi+\Psi^{\dagger}\chi^a\nabla_a 
\left(\chi^b\nabla_b\Psi\right)\right]
 \right].
\label{spin18}
\end{eqnarray}
Let us now simplify the last four terms of this equation using
symmetry arguments.  Our assumption in this case will be
$\pounds_{\xi}\Psi=0=\pounds_{\phi}\Psi$, which is of course much
stronger than the previous one made on the conserved current
1-form.

The definition of the Lie derivative of a spinor requires the
notion of Lie derivative on a fiber bundle. We refer our reader to
\cite{Godina} for a detailed discussion on this including an
exhaustive list of references. The Lie derivatives of a spinor
$\Psi$ and its adjoint $\overline{\Psi}$ along any Killing vector
field $X$ is given by
\begin{eqnarray}
\pounds_{X}\Psi=X^a\nabla_a\Psi -
\frac18\nabla_{[a}X_{b]}\gamma^a\gamma^b\Psi,\quad  
 \pounds_{X}\overline{\Psi}=X^a\nabla_a\overline{\Psi}+
\frac18\overline{\Psi}\nabla_{[a}X_{b]}\gamma^a\gamma^b.
\label{spin20}
\end{eqnarray}
It is easy to see that in a local coordinate system in which
$X^a=(\partial_x)^a$, where $x$ is the coordinate along $X^a$, the
above formula reduces to the directional partial derivative along
$X^a$. This is compatible with our common intuition about Lie
derivatives.  Thus for the customary dependence $e^{i(\omega
  t-m\phi)}$, the above conditions simply mean $\omega=0=m$. Such
condition was used previously in~\cite{Chambers:1994sz} for
spherically symmetric static spacetime.

Using Eq.s~(\ref{spin20}), (\ref{g5'n}) and (\ref{g9''}) we have
\begin{eqnarray}
\chi^a\nabla_a\Psi&=&\frac{\beta^{-1}}{4}\left[\chi_b\nabla_a\beta
  -\chi_a\nabla_b\beta\right]  
\gamma^a\gamma^b\Psi+\frac14\phi_a\nabla_b\alpha
\gamma^a\gamma^b\Psi,\nonumber\\ 
\phi^a\nabla_a\Psi&=&\frac{f^{-1}}{4}\left[\phi_b\nabla_af
  -\phi_a\nabla_bf\right] 
\gamma^a\gamma^b\Psi+\frac{f^2}{8\beta^2}
\left[\chi_a\nabla_b\alpha-\chi_b\nabla_a\alpha 
\right] \gamma^a\gamma^b\Psi.
\label{spin21}
\end{eqnarray}
The corresponding expressions for the derivatives of
$\Psi^{\dagger}$ can be found from the second of
Eq.s~(\ref{spin20}) by multiplying it by $\gamma^{(0)}$ from right
and using the anticommutation relations for the gamma matrices. We
note that since $\nabla_a\beta$, $\nabla_af$ and $\nabla_a\alpha$
are orthogonal to $\chi^a$ and $\phi^a$, in contractions like
$\chi_a(\nabla_b\beta)\gamma^a\gamma^b$,
$\phi_a(\nabla_b\alpha)\gamma^a\gamma^b$, the gamma matrices must
anticommute. Using this and Eq.~(\ref{spin2}), we find from
Eq.s~(\ref{spin21}) after a lengthy but straightforward
computation,
\begin{eqnarray}
\chi^a\nabla_a\left(\chi^b\nabla_b
  \Psi^{\dagger}\right)\Psi+\Psi^{\dagger}\chi^a\nabla_a 
\left(\chi^b\nabla_b\Psi\right)=-\frac12\left(\nabla_a\beta\right)
\left(\nabla^a\beta\right)  
\Psi^{\dagger}\Psi+\frac{f^2}{8}\left(\nabla_a\alpha\right)
\left(\nabla^a\alpha\right) 
\Psi^{\dagger}\Psi, \nonumber\\
\phi^a\nabla_a\left(\phi^b\nabla_b \Psi^{\dagger}\right)\Psi
+\Psi^{\dagger}\phi^a\nabla_a 
\left(\phi^b\nabla_b\Psi\right)=\frac12\left(\nabla_af\right)
\left(\nabla^af\right) 
\Psi^{\dagger}\Psi-\frac{f^4}{8\beta^2}\left(\nabla_a\alpha\right)
\left(\nabla^a\alpha\right) 
\Psi^{\dagger}\Psi.
\label{spin22}
\end{eqnarray}
Substituting these into Eq.~(\ref{spin18}) we get
\begin{eqnarray}
&&\overline{D}_a\left[f\beta\overline{D}^a\left(\Psi^{\dagger}
    \Psi\right)\right]=2f\beta 
\left(\overline{D}_a\Psi^{\dagger}\right)
\left(\overline{D}_a\Psi\right)-2f\beta
\left(m^2+\frac{R}{4}\right)\Psi^{\dagger}\Psi \nonumber\\ 
&&+
\frac{f\beta}{2}\left[\beta^{-2}\left(\nabla_a\beta\right)
  \left(\nabla^a\beta\right) 
 +f^{-2}\left(\nabla_af\right)\left(\nabla^af\right)
-\frac{f^2}{2\beta^2}\left(\nabla_a\alpha\right)
\left(\nabla^a\alpha\right) 
 \right]\Psi^{\dagger}\Psi, 
\label{spin23}
\end{eqnarray}
which we integrate to find
\begin{eqnarray}
\int d\Sigma\beta\left[2
\left(\overline{D}_a\Psi^{\dagger}\right)
\left(\overline{D}_a\Psi\right)
+
\frac{1}{2}\left(\beta^{-2}\left(\nabla_a\beta\right)
  \left(\nabla^a\beta\right) 
 +f^{-2}\left(\nabla_af\right)\left(\nabla^af\right)
\right.\right. \nonumber\\  
\left.\left.  
-\frac{f^2}{2\beta^2}\left(\nabla_a\alpha\right)
\left(\nabla^a\alpha\right) 
-4\left(m^2+\frac{R}{4}\right) \right)\Psi^{\dagger}\Psi\right]=0, 
\label{spin24}
\end{eqnarray}
where we have used the fact that $\displaystyle{\int f
  d\overline{\Sigma}=\frac{1}{2\pi}\int d\Sigma}$, since none of
the integrand depends on the Killing parameter $\phi$ and by
definition it ranges from $0$ to $2\pi$.  All but the fourth and
the last term in the above equation are negative definite. The
fourth term is positive and can naively be interpreted as the
repulsive effect of the spacetime rotation on matter field.  If we
set $\alpha=0$ in Eq.~(\ref{spin24}), we recover the static
spacetime equation.

We shall now examine whether the term due to rotation can dominate
the integral (\ref{spin24}).  To do this, let us consider the
Killing identity for $\phi_a$,
\begin{eqnarray}
\nabla_b\nabla^b\phi_a=-R_{a}{}^{b}\phi_b,
\label{spin25}
\end{eqnarray}
which we contract by $\phi^a$ and use Eq.~(\ref{g9''}) to get
\begin{eqnarray}
\nabla_a\nabla^af=f^{-1}\left(\nabla_af\right)
\left(\nabla^af\right)-\frac{f^3}{2\beta^2} 
\left(\nabla_a\alpha\right)\left(\nabla^a\alpha\right)
-f^{-1}R_{ab}\phi^a\phi^b. 
\label{spin26}
\end{eqnarray}
We project this equation onto $\Sigma$ using the techniques
described earlier, use Einstein's equations without backreaction
and multiply by $f^{-1}\Psi^{\dagger}\Psi$ to find
\begin{eqnarray}
D_a\left[\beta f^{-1}\Psi^{\dagger}\Psi D^af\right]
=\beta\left[\left(-\frac{f^2}{2\beta^2} 
\left(\nabla_a\alpha\right)\left(\nabla^a\alpha\right)
+\Lambda\right) \Psi^{\dagger}\Psi 
+f^{-1}\left(D_a\left(\Psi^{\dagger}\Psi\right)\right)
\left(D^af\right) \right],
\label{spin27}
\end{eqnarray}
which we integrate between the two horizons. The boundary integrals
go away and we combine the vanishing volume integral with
Eq.~(\ref{spin24}) to get

\begin{eqnarray}
\int d\Sigma\beta\left[
\left(\overline{D}_a\Psi^{\dagger}\right)
\left(\overline{D}_a\Psi\right)
+
\frac{1}{2}\left(\beta^{-2}\left(\nabla_a\beta\right)
  \left(\nabla^a\beta\right) 
 +\frac{f^{-2}}{2}\left(\nabla_af\right)\left(\nabla^af\right)
-4\left(m^2+\frac54\Lambda\right) \right)
\Psi^{\dagger}\Psi\right. \nonumber\\ 
\left.
+\left(\overline{D}_a\Psi-
  \frac{f^{-1}}{2}\Psi\overline{D}_af\right)^{\dagger} 
\left(\overline{D}^a\Psi- \frac{f^{-1}}{2}\Psi \overline{D}^af\right)
\right]=0,
\label{spin28}
\end{eqnarray}
where we have used the fact that $\nabla_af=D_af=\overline{D}_af$
(Eq.s~(\ref{g5'nn})).  All the terms are negative definite now,
which shows that $\Psi=0$ throughout our region of interest, which
is the desired no hair result. This result clearly holds for
$\Lambda=0$ provided we impose suitable fall-off condition at
spatial infinity.  This also holds for an asymptotically anti-de
Sitter spacetime if in addition to the fall-off condition, we
assume that $\displaystyle{m^2\geq
  \frac54\left\vert\Lambda\right\vert}$, which means that the
Compton wavelength of the spinor is small compared to the AdS
length scale.

\section{Summary}
In this work we have proved no hair theorems for massive spin-$2$
and spin-$\frac12$ fields for general stationary axisymmetric de
Sitter black hole spacetimes. The existence of quantum hair for the
spin-$2$ field was also discussed. Since spinors do not satisfy any
classical energy condition, the no spinor hair could only be proved
upon imposition of weakness condition. The backreaction of spinors
should involve renormalization of the energy-momentum tensor, which
seems an interesting problem in stationary axisymmetric spacetime.
It will be interesting to investigate the situation when the spinor
gets coupled to a gauge field, a Maxwell field for example.


\section*{Acknowledgment}
SB thanks A. Basu for useful discussions.

\vskip 1cm

\end{document}